\documentclass[aip,reprint]{revtex4-1}
\usepackage{graphicx}% Include figure files
\usepackage{dcolumn}% Align table columns on decimal point
\usepackage{bm}% bold math
\usepackage[utf8]{inputenc}
\usepackage[T1]{fontenc}
\usepackage{mathptmx}
\usepackage{etoolbox}
\usepackage{siunitx}
\usepackage{float}
\usepackage{subfig}
\usepackage{comment}

\draft % marks overfull lines with a black rule on the right
\usepackage[colorlinks=true, allcolors=blue]{hyperref}

\begin{document}

%\preprint{}

\title{Experimental Evidence of Amplitude-Dependent Surface Wave Dispersion via Nonlinear Contact Resonances} %Title of paper



\author{Setare Hajarolasvadi}
%\thanks{}
\affiliation{Department of Civil and Environmental Engineering, University of Illinois at Urbana-Champaign, Urbana, IL 61801, USA}

\author{Paolo Celli}
\affiliation{Department of Civil Engineering, Stony Brook University, Stony Brook, NY 11794, USA}

\author{Brian Kim}
\affiliation{Division of Engineering and Applied Science, California Institute of Technology, Pasadena, CA 91125, USA}

\author{Ahmed E. Elbanna}
\affiliation{Department of Civil and Environmental Engineering, University of Illinois at Urbana-Champaign, Urbana, IL 61801, USA}

\author{Chiara Daraio}
\email[To whom correspondence should be addressed: ]{daraio@caltech.edu}
\affiliation{Division of Engineering and Applied Science, California Institute of Technology, Pasadena, CA 91125, USA}


\begin{abstract}
In this letter, we provide an experimental demonstration of amplitude-dependent dispersion tuning of surface acoustic waves interacting with nonlinear resonators. Leveraging the similarity between the dispersion properties of plate edge waves and surface waves propagating in a semi-infinite medium, we use a setup consisting of a plate with a periodic arrangement of bead-magnet resonators along one of its edges. Nonlinear contact between the ferromagnetic beads and magnets is exploited to realize nonlinear local resonance effects. First, we experimentally demonstrate the nonlinear softening nature and amplitude-dependent dynamics of a single bead-magnet resonator on both rigid and compliant substrates. Next, the dispersion properties of the system in the linear regime are investigated. Finally, we demonstrate how the interplay of nonlinear local resonances with plate edge waves gives rise to amplitude-dependent dispersion properties. The findings will inform the design of more versatile surface acoustic wave devices that can passively adapt to loading conditions.
\end{abstract}

\pacs{}% insert suggested PACS numbers in braces on next line

\maketitle %\maketitle must follow title, authors, abstract and \pacs

Surface acoustic waves (SAWs) have broad applications in science and engineering. At the micro and nano scales, these waves are of interest in the design of radio-frequency filters for wireless telecommunication systems \cite{Benchabane2019} as well as biosensors for medical diagnostics \cite{Lange2008}. At larger scales, the study of these waves is essential for protecting the built environment from the damaging effects of seismic waves \cite{Brule2013, Muhammad2019}. The advent of metamaterials has realized unique engineering solutions for manipulating these waves at vastly different frequencies. For example, phononic crystals in the form of architected surface layers have been used to design SAW filters, space-saving reflective gratings and waveguides \cite{Wu2015}. Periodic arrangements of local resonators have also been used to achieve subwavelength wave filtering and waveguiding \cite{Khelif2010}, as well as high-resolution imaging \cite{Addouche2014, Fuentes-Dominguez2021}. 

Once fabricated, metamaterials for SAW control are usually bound to produce a desired effect at a specific frequency range. Recently, efforts have been undertaken to increase the versatility of these systems by making their response tunable \cite{Cselyuszka2016, Palermo2019}, or non-reciprocal~\cite{Palermo2020,wu2021non}. Most previous work on SAW tunability has focused on using external stimuli (e.g., thermal, magnetic, electrical) for tuning the wave-control capabilities of these systems \cite{Palermo2020, wu2021non}. However, the addition of external stimuli adds complexity to the system. A desirable alternative is the design of self-tunable SAW devices, which can \emph{passively adapt} to the loading conditions without the need for external stimuli. Incorporating nonlinearity in the design of metamaterials provides an opportunity to explore amplitude-dependent self-tuning for SAWs.  

Nonlinear metamaterials offer enhanced control over wave transmission compared to their linear counterparts. Several exotic features have already been demonstrated in these systems, including self-tunability \cite{Narisetti2010, Jiao2019}, nonreciprocity \cite{Mojahed2019, Moore2018, Fronk2019}, energy tunneling and localization \cite{Jiao2018} and, more recently, the emergence of subharmonic bandgaps \cite{Zega2020}.  Theoretical frameworks have been developed for determining the dispersion properties of nonlinear phononic lattices/crystals \cite{Narisetti2012, Manktelow2013}. More recent theoretical investigations have focused on the effects of material or geometric nonlinearity in elastic metamaterials \cite{Khajehtourian2014}. The effects of nonlinear local resonators, on the other hand, have been mostly studied in the context of discrete systems \cite{Lazarov2007, Manktelow2017}. Limited works exist on wave propagation in systems consisting of an array of nonlinear resonators embedded in linear elastic continua \cite{Fang2018}. Even though the interaction of SAWs with contact-based resonators, with an inherently nonlinear nature, has been studied previously, such studies are based on the assumption that the amplitude of the propagating waves is small and that the nonlinear stiffness can be linearized 
 \cite{Boechler2013,Wallen2017}. 

Amplitude-dependent resonance is a well-documented phenomenon in nonlinear dynamics \cite{Nayfeh1995}. Several works have documented this effect for Hertzian contact resonators \cite{Rigaud2003, Perret-Liaudet2003}. A recent study on a cylindrical rod in contact with a bead provides experimental proof that the nonlinear properties of the contact lead to amplitude-dependent resonance shifts \cite{Merkel2019}. A notable experimental work had previously demonstrated how resonance shifts in a one-dimensional chain of beads connected with nonlinear springs are intimately related to shifts in dispersion curves for the overall system \cite{Manktelow2014}. In a more recent and relevant work, the propagation of Rayleigh waves in a half-space coupled to nonlinear resonators was considered. The authors provided a theoretical description of Rayleigh wave dispersion in the presence of hardening and softening interaction forces, and validated their findings using Finite Element simulations \cite{Palermo2022}. However, experimental investigations of nonlinear dispersion shifts for SAWs have remained unexplored.

\begin{figure}[!htb]
    \includegraphics[width=1\columnwidth]{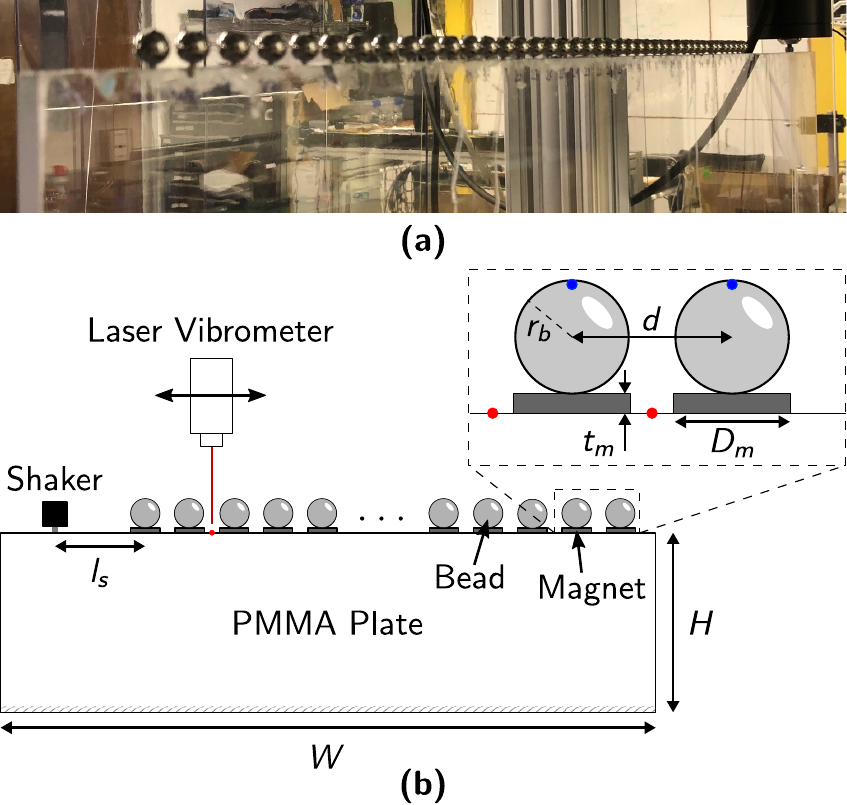}
    \caption{\textcolor{black}{Configuration of the table-top experimental setup. (a) Image of the setup, and (b) its schematics showing the components' dimensions as well as the location of observation points on the plate's edge (red) and on the resonators (blue).} }
    \label{fig: tabletop-config}
\end{figure}

In this work, we leverage an experimental setup similar to that of Ref.~\cite{Palermo2019} and exploit the nonlinear dynamics of an array of contact resonators to achieve amplitude-dependent dispersion properties for plate edge waves. The compact tabletop experimental setup is shown in Fig.~\ref{fig: tabletop-config}. 
%%%%%
%%%%%
It consists of an acrylic plate of dimensions $608\times912\times8$ \si{\milli\meter} ($H\times W \times t$), Young's modulus $E=5.5$ \si{\giga\pascal}, Poisson ratio $\nu=0.35$, and density $\rho=1190$ \si{\kilo\gram\per\cubic\meter}. The plate is clamped to an optical table at the bottom along the longer edge. A set of 41 disk magnets (K\&J magnetics DH101; NdFeB, Grade N42) are glued at equal distances of $d=15$ \si{\milli\meter} on its top edge. The magnets have a diameter of $D_m=2.5$ \si{\milli\meter} and a thickness of $t_m=0.8$ \si{\milli\meter}. \textcolor{black}{The Young's modulus $E_m$ and Poisson ratio $\nu_m$ of the magnet are 190 \si{\giga\pascal} and 0.3, respectively \cite{Ashby2018}.} Steel beads (McMaster-Carr 9642K49) with radius $r_b=4.8$ \si{\milli\meter} and mass $m_b=3.5$ \si{\gram} are placed on top of each magnet. \textcolor{black}{The steel beads have a Young's modulus of $E_b=210$ \si{\giga\pascal} and a Poisson ratio of $\nu_b=0.3$ \cite{Ashby2018}.}
The bead-magnet assemblies will serve as nonlinear mechanical oscillators. A vibration exciter (HBK Type 4810) is glued to the plate at a distance of $l_s=168$  \si{\milli\meter} from the first bead. A signal generator (Agilent 33220A) and power amplifier (HBK Type 2718) are used to drive the shaker and excite vertically-polarized edge waves along the edge of the plate. A laser doppler vibrometer (LDV, Polytec OFV-5000) is mounted on a linear stage and a motor is used to move the vibrometer and consecutively measure the vertical velocity component at desired observation points. Measurement data is acquired using an oscilloscope (Tektronix DPO3034). 

We use the analytical dispersion relation for thin semi-infinite plates with \textcolor{red}{stress-free} boundary conditions \cite{Wilde2019} to predict the phase velocity $c_R$ of edge waves in a pristine plate:
	%%%%%%%%%%%%%%%%%%%%%%%
	\begin{equation}\label{eq: semi-infinite-plate-dispersion}
	\left(2-\frac{c_R^2}{c_T^2}\right)^2-4\sqrt{\left(1-\frac{c_R^2}{c_T^2}\right)\left(1-\frac{c_R^2}{v_P^2}\right)}=0,
	\end{equation}
	%%%%%%%%%%%%%%%%%%%%%%%
	where $v_p=\left[E/\rho(1-\nu^2)\right]^{1/2}$ is the velocity of dilational waves in a thin plate and $c_T=\left[E/2\rho(1+\nu)\right]^{1/2}$ is the shear wave speed \cite{Wilde2019}. We note that this equation is similar to the one describing the dispersion relation of Rayleigh waves in a half-space. Using this equation, we make a theoretical prediction of $c_R=1205$ \si{\meter\per\second} for the phase velocity of edge waves propagating on a pristine plate.
 
\begin{figure}[!htb]
    \centering
    \includegraphics[width=1\columnwidth]{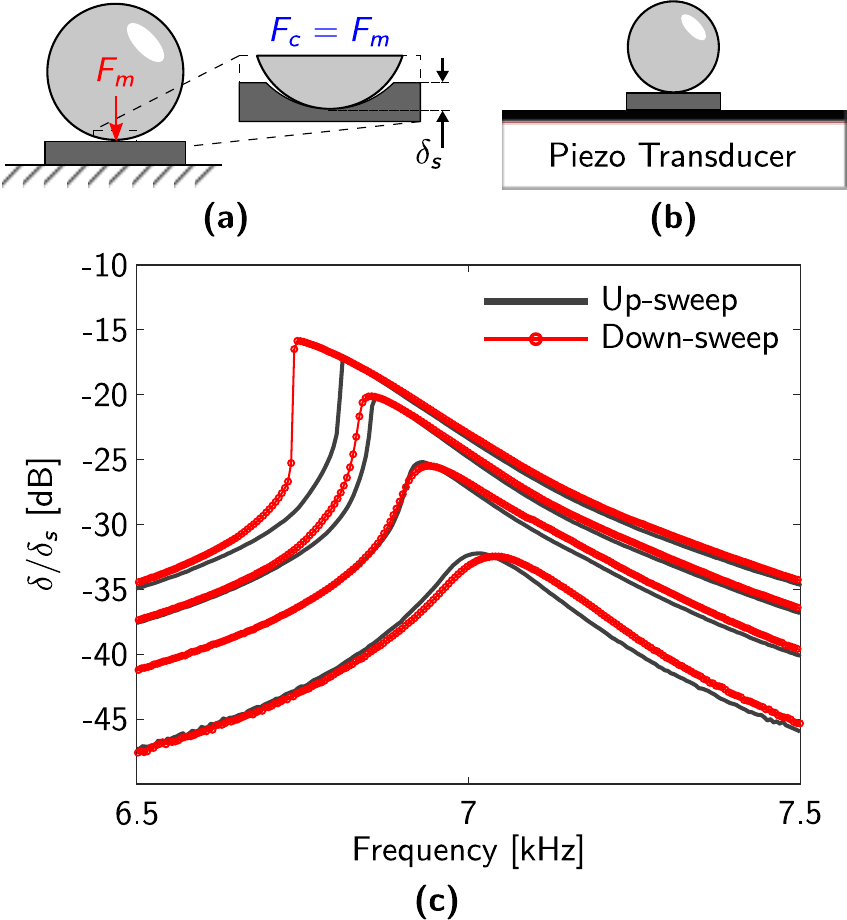}
    \caption{\textcolor{black}{Single bead-magnet assembly on a rigid substrate. (a) Schematics of the problem and the forces exerted on the bead. The insert shows the static overlap $\delta_s$ between the surfaces in contact at rest. $F_c$ and $F_m$ are the contact and magnetic force, respectively. (b) Schematic of the experimental setup. (c) Up-sweep (black) and down-sweep (red) experimental frequency response functions for the bead-magnet resonator at 8 \si{\milli\volt}-interval excitation amplitudes.}}
    \label{fig: single}
\end{figure}

 Prior to investigating the dynamics of the overall system, we characterize a single bead-magnet oscillator. We start by experimentally investigating the resonance characteristics of a single bead-magnet assembly on a rigid substrate. To do so, we attach the disk magnet to the surface of a piezoelectric transducer (Panametrics-NDT V1011) using cyanoacrylate glue. The bead is then placed on top of the magnet \textcolor{black}{(Fig.~\ref{fig: single} (a), (b))}. 
 %%%%%%%%%%%%%%%
%%%%%%%%%%%%%%%
 Due to the importance of the contact surface in these experiments, we thoroughly clean the surface of the magnet as well as the steel bead before they get in contact. A Stanford SR 860 analyzer is used for the excitation in a sine sweep mode from five to eight \si{\kilo\hertz}, and the laser Doppler vibrometer is used for measuring the velocity response of the bead. We start at an amplitude of 8 \si{\milli\volt} and repeat the test by increasing the excitation amplitude at 8 \si{\milli\volt} intervals. Fig.~\ref{fig: single} (c) shows the frequency response plots of the bead-magnet resonator. The steady-state amplitudes have been normalized by the static overlap between the bead and the magnet $\delta_s$. The black and red curves show results for sweep-up and sweep-down tests, respectively. 

 At low excitation amplitudes, the bead-magnet assembly is expected to behave as a linear oscillator. Thus, the frequency response curves from up and down sweeps coincide. The underlying linear natural frequency of the oscillator is approximately $f_r=7$ \si{\kilo\hertz}. The linearized normal stiffness of the resonator $k_N$ can then be estimated as $m_b(2\pi f_r)^2$. Assuming a Hertzian contact law between the bead and the magnet, the linearized normal stiffness may also be written in terms of the static overlap $\delta_s$ as $k_N=2E^*r_b^{1/2}\delta_s^{1/2}$ \cite{Palermo2019}, where $E^*=\left[(1-\nu_b^2)/E_b+(1-\nu_m^2)/E_m\right]^{-1}$. From here, the static overlap $\delta_s$ is approximately determined as 200 \si{\nano\meter}.
 %{\color{red} The linearized normal stiffness of the resonator $k_N$ can then be estimated as $m_b(2\pi f_r)^2$. Assuming a Hertzian contact law between the bead and the magnet, the linearized normal stiffness may also be written in terms of the static overlap $\delta_s$ as $k_N=2E^*r_b^{1/2}\delta_s^{1/2}$ \cite{Palermo2019}, where $E^*=\left[(1-\nu_b^2)/E_b+(1-\nu_m^2)/E_m\right]^{-1}$. From here, the static overlap $\delta_s$ is determined as ~200 \si{\nano\meter}}.

As the excitation amplitude is increased, nonlinearity bends the frequency response and shifts the locus of the peak amplitude to lower frequencies. This is characteristic of a softening nonlinear response. In addition, the differences between up and down swept curves become increasingly stark. Emergence of jumps in the frequency response at an excitation amplitude of 32 \si{\milli\volt} indicates the resonator's loss of stability and is another evidence of the inherent nonlinear properties of contact resonance. 

Next, we study the dynamics of a single bead-magnet oscillator on an acrylic plate. We present experimental evidence of higher harmonic generation and resonance frequency shifts in the contact resonator's response. The schematic of the experimental setup is similar to the one shown in Fig. ~\ref{fig: tabletop-config}, with the difference that all bead magnet resonators, except the one closest to the shaker, are removed. %\textcolor{red}{The schematic of the experimental setup is similar to the one shown in Fig. ~\ref{fig: tabletop-config}, with the difference that all bead magnet resonators, except the one closest to the shaker, are removed.} 
The linear resonance frequency of the oscillator is identified at roughly 5.15 \si{\kilo\hertz} using a broadband excitation. This shows a shift of approximately 1.85 \si{\kilo\hertz} in comparison to measurements on a rigid substrate (Fig. \ref{fig: single}), which can be attributed to the substrate's compliance and its coupling to the rigid contact dynamics. A similar effect has been reported in previous work \cite{Palermo2019}. Based on the determined resonance frequency in the linear regime, we use a narrow-band, slow \textcolor{black}{(200 \si{\second^{-2}})} sweep-down excitation from 6 \si{\kilo\hertz} to 4 \si{\kilo\hertz} to characterize the nonlinear response of the oscillator. The voltage was set to 100 \si{\milli\volt}, and three different excitation amplitudes (10, 20, and 30 \si{\deci\bel}) were chosen by changing the gain on the amplifier. The response of the bead was directly recorded by the LDV. 
%%%%
\begin{figure}[!htb]
	\centering
	\includegraphics[width=1.0\columnwidth]{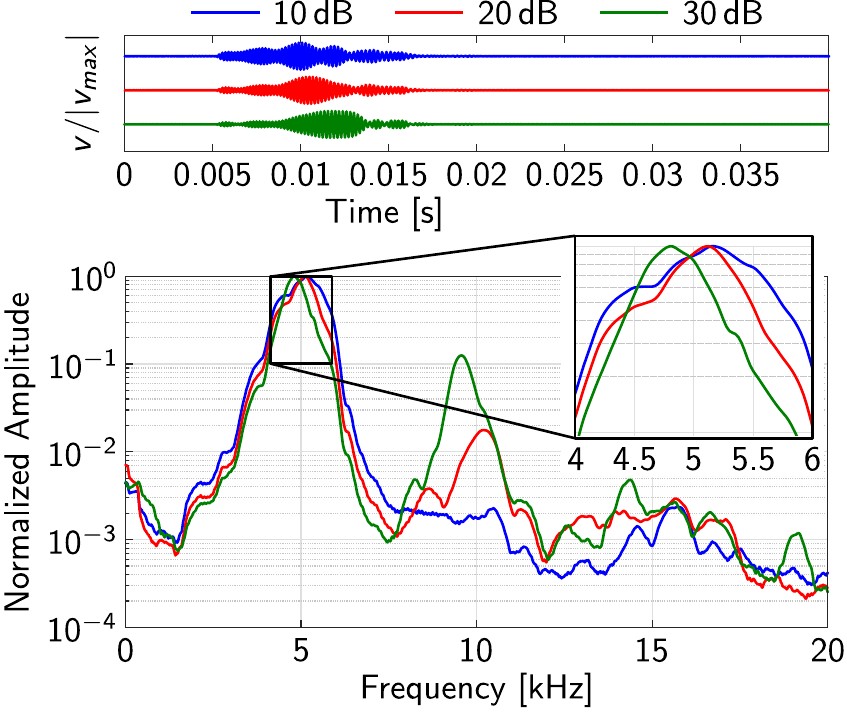}
	\caption{Experimental results for the single bead-magnet resonator on the acrylic plate. Top panel shows the time history responses at the three gain levels, shifted for better visualization. The bottom panel shows the frequency domain response of the single bead-magnet resonator.}
    \label{fig: single-resonator-on-plate-th-fd}
\end{figure}
%%%%
Fig.~\ref{fig: single-resonator-on-plate-th-fd} shows the time history as well as frequency spectrum for the bead's response. A moving average filter was used to postprocess the response. Table~\ref{tab: single-bead-on-plate-resonance} summarizes the main peaks in the frequency spectrum for different levels of gain. The tabular data shows that the primary resonance frequency $f_1$ shifts to lower frequencies as the amplitude of the excitation increases. As discussed in the previous section, this is characteristic of a softening nonlinear behavior. Additionally, increasing the excitation amplitude leads to the generation of a higher harmonic $f_2$ at twice the resonance frequency.
%%%%
\begin{table}[!htb]
    \centering
    \begin{tabular}{c c c c}
    \hline
    Gain [dB] & $f_1$ [kHz] & $f_2$ [kHz] & $f_2/f_1$\\
    \hline\hline
    10 & 5.15    & - & -\\
    20 & 5.12 & 10.2 & 1.99\\
    30 & 4.80 & 9.57 & 1.99\\
    \hline
    \end{tabular}
    \caption{Resonance frequencies of the single bead-magnet resonator on the acrylic plate.}
    \label{tab: single-bead-on-plate-resonance}
\end{table}
%%%%
Another interesting feature is observed in the results by comparing the time history plots (top panel in Fig.~\ref{fig: single-resonator-on-plate-th-fd}). \textcolor{black}{The velocity of the bead $v$ has been normalized by peak velocity $v_{max}$ in each case.} At lower gains, the rise and fall of the amplitude at resonance is symmetric in shape. However, at 30 \si{\deci\bel} gain, the descent from resonance is abrupt, suggesting loss of stability, a feature common to nonlinear resonance phenomenon.

We now move on to determine the \textcolor{black}{dispersion properties of edge waves for a plate with a periodic array of bead-magnet resonators.} The experimental setup is shown in Fig.~\ref{fig: tabletop-config}. 
%%%%%
\begin{figure*}[!htb]
    \includegraphics[width=1.0\textwidth]{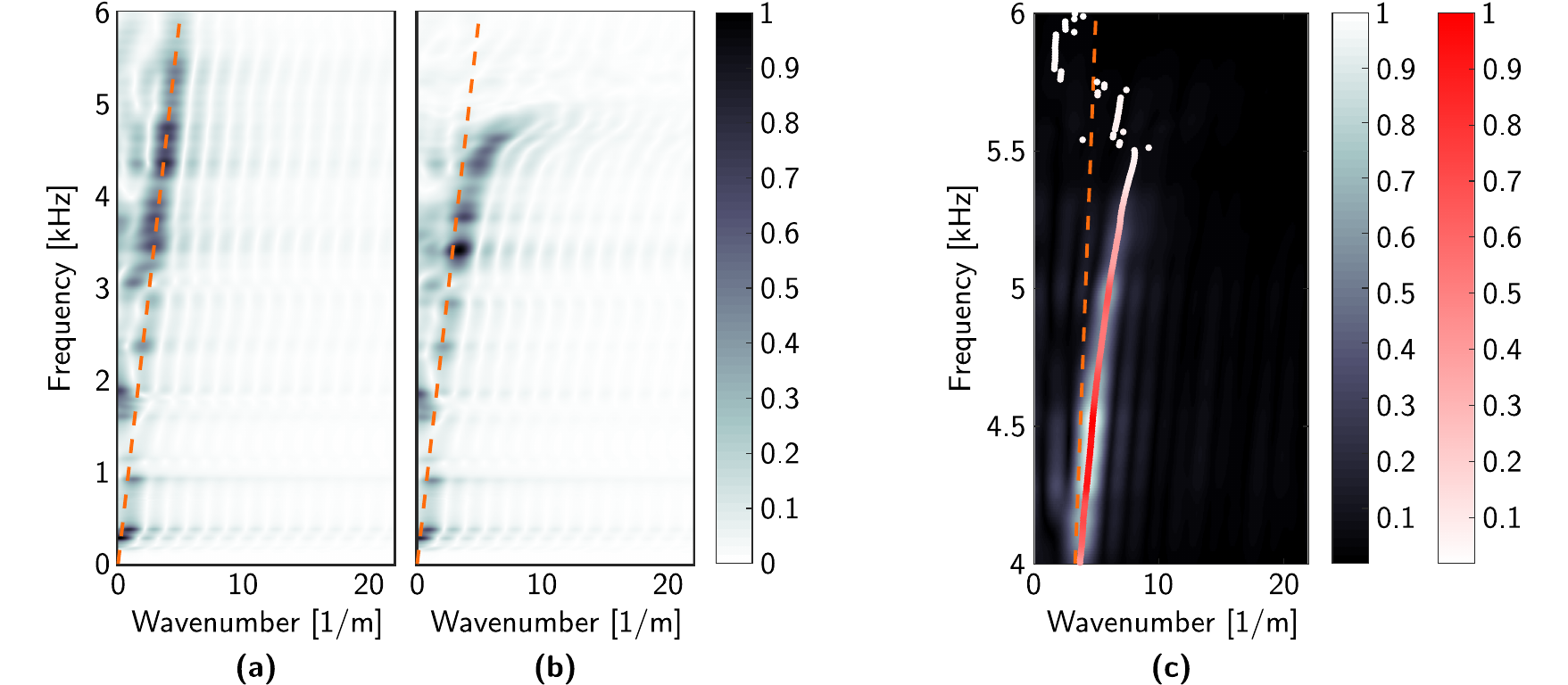}
    \caption{Dispersion reconstruction using experimental measurements: (a) the pristine plate in the linear regime, (b) the plate with an array of bead-magnet resonators in the linear regime (10 \si{\decibel} gain), and (c) the plate with an array of bead-magnet resonators in the nonlinear regime (20 \si{\decibel} gain). The dashed orange line indicates the Rayleigh wave dispersion curve.}
    \label{fig: dispersion reconstruction}
\end{figure*}
%%%%%
Two primary modes of excitation are utilized in this experiment: a wide-band sweep at low amplitudes that captures the linear response of the system, and a narrow-band slow sweep at higher amplitudes that is used to investigate the nonlinear characteristics of the system. Dispersion reconstruction  in the linear regime is carried out using a wide-band fast \textcolor{black}{(590 \si{\second^{-2}})} sweep-up excitation from 100 \si{\hertz} to 6 \si{\kilo\hertz}. Due to evidence of softening nonlinearity in the response of the oscillator, the nonlinear system response is best characterized using a narrow-band slow \textcolor{black}{(200 \si{\second^{-2}})} sweep-down excitation from 6 \si{\kilo\hertz} to 4 \si{\kilo\hertz}. For all the above cases, we study the interaction of surface waves with the array of bead-magnet resonators by recording the vertical velocity at 42 stations on the edge of the plate, marked as red circles in Fig.~\ref{fig: tabletop-config}. The distance between adjacent observation points is $15$ \si{\milli\meter}. 

To unveil the linear response of the system, which is used as a baseline to understand the effects of nonlinearity, we reconstruct the dispersion for the pristine plate, and the plate with an array of bead-magnet resonators. Fig.~\ref{fig: dispersion reconstruction} (a) shows the experimental dispersion curve for the pristine plate in a gray-scale contour. The broadband chirp generated by the shaker travels along the plate's edge dispersionless, as expected. The dashed orange curve shows the Rayleigh wave speed $c_R=1205$ \si{\meter/\second} given by Eq.~\ref{eq: semi-infinite-plate-dispersion}. We note that gluing the array of magnets to the plate's edge does not introduce any dispersion effects since the magnets' mass ($\approx 0.03$ \si{\gram}) is negligible. Fig.~\ref{fig: dispersion reconstruction} (b) shows the dispersion curves for the plate with an array of bead-magnet resonators. Placing the contact resonators on the plate's edge leads to hybridization between the traveling wave and the resonance modes. The slow-propagating flat branch observed in the dispersion plot is a result of SAW interaction with vertical resonances of the bead-magnet resonators~\cite{Wallen2017, Palermo2019}. The frequency where the branch flattens agrees well with the primary resonance frequency for a single bead on the acrylic plate, determined previously.

In order to investigate the behavior of the system in the nonlinear regime, we use the slow narrow-band chirp. Three different excitation amplitudes were chosen by changing the gain on the amplifier. In order to quantify confidence in the experimental results, three sets of measurements were done at each gain, leading to nine sets of data in total. After each measurement, all beads were removed, cleaned and placed on the magnets again. This was done to ensure that the results were not significantly affected by the uncertainties associated with the bead-magnet contact surface. Furthermore, the order in which the nine experiments were done was random. For each set of measurements at constant amplitude, recorded spatio-temporal data on the plate's edge was postprocessed using 2D Fourier transforms. The average of normalized Fourier amplitudes over each three sets of measurements was then used to visualize the system's dispersion.

Fig.~\ref{fig: dispersion reconstruction} (c) shows the reconstructed dispersion for the structure at 20 \si{\decibel} gain. The gray-scale contour shows the full 2D visualization of response in the wave number-frequency domain, with white showing the highest intensity. At each discrete frequency value, the wave-number corresponding to the maximum Fourier amplitude was identified. This gives the overlaid scattered plot in a gradient of red. The color of these markers at each point indicates the intensity of the normalized Fourier amplitude, with white having the lowest intensity and red the highest. This approach will prove itself crucial later for comparing the dispersion branches at different gain levels. It also helps highlight data points of greater significance. For example, we can see that the data points lying outside the sound cone are of extremely low intensity and therefore of little significance. Thus, we can safely ignore them. The band structure at the other two excitation amplitudes (10 and \textcolor{black}{30} \si{\decibel}) is constructed in a similar manner.

\begin{figure}[!htb]
    \centering
    \includegraphics[width=1.0\columnwidth]{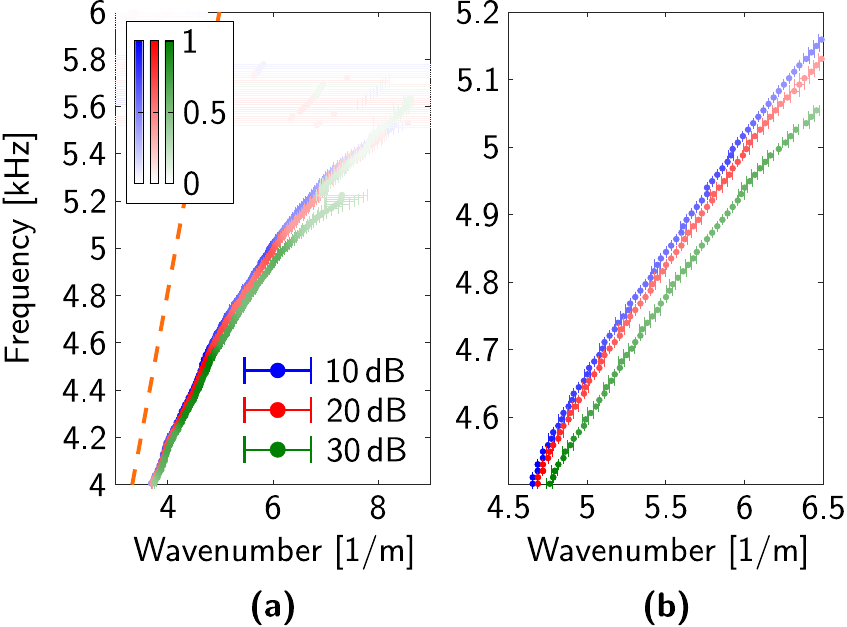}
    \caption{Dispersion reconstruction in a limited frequency region: (a) overlaid dispersion branches for the three excitation levels. (b) same as (a), zoomed in a more limited region to highlight the dispersion shift. The colored markers are graded according to the magnitude of the normalized Fourier amplitude.}
    \label{fig: nlr-dispersion-overlaid}
\end{figure}

Fig.~\ref{fig: nlr-dispersion-overlaid} (a) shows the dispersion branches reconstructed at the three different amplitudes. 
%%%%
%%%%
The scattered plots are now shown in the form of error-bar plots; that is, at each frequency, the marker indicates the mean and the horizontal bar shows the standard deviation of the wavenumber corresponding to the maximum Fourier amplitude for the three sets of measurements. It is clear that in the regions where Fourier amplitude is high, standard deviation is extremely small. On the contrary, as intensity approaches zero, the standard deviation becomes very large. In the regions of high intensity and low standard deviation, the figure shows that increasing the excitation amplitude shifts the dispersion curve to lower frequencies. In other words, with an increase in the excitation amplitude, the wavenumber corresponding to a fixed frequency increases. This is quantitatively shown in Table \ref{tab: dispersion shift} for three select frequencies.   
%%%%
\begin{table}[!htb]
    \centering
    \begin{tabular}{c c c c}
    \hline
    & \multicolumn{3}{c}{Wavenumber [1/\si{\meter}]}\\
    \hline\hline
     Frequency [\si{\kilo\hertz}] & 10 \si{\decibel} & 20 \si{\decibel} & 30 \si{\decibel}\\
    \hline
    4.2 & 4.65 & 4.69 & 4.76\\
    4.7 & 5.1 & 5.14 & 5.28\\
    5.2 & 6.63 & 6.76 & 7.28\\
    \hline
    \end{tabular}
    \caption{Increase of the wavenumber value with an increase in excitation amplitude at sample low, medium and high frequencies.}
    \label{tab: dispersion shift}
\end{table}
%%%%

Fig.~\ref{fig: nlr-dispersion-overlaid} (b) shows the dispersion branches in a more limited wave number-frequency region to highlight the amplitude-dependent dispersion shift. Just like in the softening theoretical prediction of Palermo et al.\cite{Palermo2022}, we observe that: i) an increase in excitation amplitude in the presence of softening nonlinearity shifts the dispersion branch down; and ii) higher-amplitude branches tend to terminate early, i.e., at lower wavenumbers compared to their low-amplitude counterparts. The early termination may be explained by the onset of instability for the surface resonators \cite{Palermo2022}.  Comparing wavenumber-frequency pairs at a certain threshold of the normalized Fourier amplitude intensity proves useful for quantifying the early termination of dispersion branches. For example, a 0.3 normalized Fourier amplitude intensity corresponds to the point (7.0204 1/\si{\meter}, 5302.4 \si{\hertz}) on the  dispersion branch at 10 \si{\decibel} excitation amplitude. However, at the same intensity, wavenumber-frequency pairs at 20, and 30  \si{\decibel} are (6.901 1/\si{\meter}, 5235.7 \si{\hertz}) and (6.6515 1/\si{\meter}, 5092.6 \si{\hertz}), respectively. 
%The data shows a frequency shift of  66.7 Hz (~1.26\%) from 10 \si{\decibel} to 20 \si{\decibel} and 143.1 Hz (~2.73\%) from 20 \si{\decibel} to 30 \si{\decibel}. 
As such, the termination wavenumber decreases by ~1.7\% from 10 \si{\decibel} to 20 \si{\decibel} and 3.62\% from 20 \si{\decibel} to 30 \si{\decibel}. 

In conclusion, we have investigated the interaction of surface acoustic waves with nonlinear contact resonators and provided experimental evidence of amplitude-dependent surface wave dispersion. Careful investigation of the bead-magnet's dynamics revealed several features common to softening nonlinear oscillators, such as amplitude-dependent resonance frequency, higher harmonic generation and loss of stability. These characteristics make the array of bead-magnet assemblies suitable for use as nonlinear resonators in a compact setup. In the current setup, the surface wave energy available for interaction with the nonlinear resonators is limited due to the maximum force rating of the shaker and the overall energy loss in the system. This prevents the realization of more significant shifts, such as those induced by external stimuli \cite{Palermo2019}. However, this proof-of-concept demonstration serves as a motivation for other researchers to design novel solutions to induce more dramatic self-tuning effects for SAWs. These could include creating a waveguide close to the plate's edge to maximize the surface wave energy and exploring the dynamics of the resonators in the vibroimpact region. Loss of contact nonlinearity for Hertzian contact resonators has been shown to induce more significant resonance frequency shifts \cite{Rigaud2003,Perret-Liaudet2003}.  

\begin{acknowledgments}
 We thank Professor Alexander Vakakis and fellow researchers Alireza Mojahed, Joaqin Garcia-Suarez, and Danilo Kusanovic for the stimulating discussions. This research has been supported by the US National Science Foundation Grant EFRI-1741565, the National Science Foundation Career Award No. 1753249, and the Graduate College Dissertation Completion Fellowship Award provided by the University of Illinois at Urbana-Champaign. 
\end{acknowledgments}
%%%%
%\label{}
%\subsection{}
%\subsubsection{}

% If in two-column mode, this environment will change to single-column format so that long equations can be displayed. 
%\begin{widetext}
%$$\mbox{put long equation here}$$
%\end{widetext}

% Figures should be put into the text as floats. 
% Use the graphics or graphicx packages (distributed with LaTeX2e).
% Put the label that you will use with \ref{} command in the braces of the \label{} command.
%
% \begin{figure}
% \end{figure}

% \begin{table}
% \end{tabular}
% \end{table}

%\begin{acknowledgments}
% Put your acknowledgments here.
%\end{acknowledgments}

\nocite{*}

\bibliography{nlr_saw}

\providecommand{\noopsort}[1]{}\providecommand{\singleletter}[1]{#1}%

\begin{comment}
    %

\end{comment}

\end{document}